\documentclass[10pt,twoside]{article}
\usepackage {FIE}
\usepackage {epsfig}
\usepackage{fancyhdr}
\newenvironment{mylist}{%
\begin{list}{$\bullet$}{\setlength\itemsep{0\baselineskip}%
}%
}{%
\end{list}%
}

\begin{document}
\title{Clown: a Microprocessor Simulator for Operating System
  Studies}
\author{Dmitry~Zinoviev\\
  Computer Science Department, Suffolk University\\ 
  32 Derne St., Boston, MA, 02114 USA\\
  Dmitry@mcs.suffolk.edu}
\date{}
\maketitle
\setcounter{page}{99}
\pagestyle{fancy}

\noindent%
\vskip\baselineskip

\abstract{%
In this paper, I present the design and implementation of Clown --- a
simulator of a microprocessor-based computer system specifically
optimized for teaching operating system courses at undergraduate or
graduate levels. The package includes the simulator itself, as well as
a collection of basic I/O devices, an assembler, a linker, and a disk
formatter. The simulator architecturally resembles mainstream
microprocessors from the  Intel 80386 family, but is much easier to
learn and program. The simulator is fast enough to be used as an
emulator --- in the direct user interaction mode.
}

\section{A NEED FOR A SIMULATOR}

An important part of the agenda of a college-level operating system
course is to examine the interaction between an operating system and
computer hardware. Assembly programming teaches students to think
logically, waste no byte and no CPU cycle. Knowing the hardware helps
students to understand the operation of such foundational mechanisms
as memory protection, process dispatching, input/output, and file
system organization. It also makes it clearer the motivations behind
certain OS design decisions.  Last, but not least, from the practical
point of view, exposing students to low-level programming prepares
them for potential projects involving embedded systems and hand-held
devices.

Elements of low-level assembly programming can be also found in
computer architecture courses. Many universities\footnote{Suffolk
University being one of them.} continue to offer general
assembly programming courses where students learn how to extract the
ultimate performance from the computer hardware.

Traditionally, colleges have been using various RISC architectures
(such as MIPS or RS6000) or Motorola 68x family as their primary
hardware platforms. RISC cores are reasonably simple and
regular. However, this trend seems to be rapidly disappearing in favor
of the industrial mainstream Intel32 architecture. It should be also
noted that from the OS development point of view, RISC cores lack many
important features, such as segmentation (for superior memory
protection) and non-trap-based system call support.

On the other hand, Intel32 CISC architecture is hard to learn. The
instruction set is redundant, and the instruction format is highly
irregular. This makes Intel32 system programming challenging,
especially for undergraduate students. A need clearly exists for a
good microprocessor simulator that could be used in an OS course (and
possibly in other related courses).

\section{EXISTING SIMULATORS}
Many microprocessor simulators have been developed, but most of them
do not address the topic from the OS study point of view.

Some of them simulate RISC or otherwise ``inappropriate'' targets
(e.g., Ant-32~\cite{ellard02}, MicSim~\cite{merz96}, Microprocessor Trainer
Simulator~\cite{caldwell95}, and various Intel 8085 simulators, such
as described in~\cite{insoluz98}).

Other simulators are too detailed (such as VMware~\cite{vmware} and
SID~\cite{sid01}). They are simulating the computer hardware as close
as possible, thus defeating the whole purpose of using a simulator in
an undergraduate-level class. On the other hand, many simulators
designed for educational purposes, are oversimplified
(MSFB~\cite{bauers}, also \cite{hill94functional} and
\cite{caldwell95}). Being good for an introductory computer hardware
course, they fail to provide substantial mechanisms for building
advanced operating systems.

To summarize, existing simulators are either optimized to be used in
industry or in a hardware-oriented course~\cite{patterson96}, but not
in a ``classic'' OS course~\cite{silberschatz02}, or they are
intentionally hiding hardware from the upper OS layers.

\begin{figure*}[t!]\centering
\epsfig{file=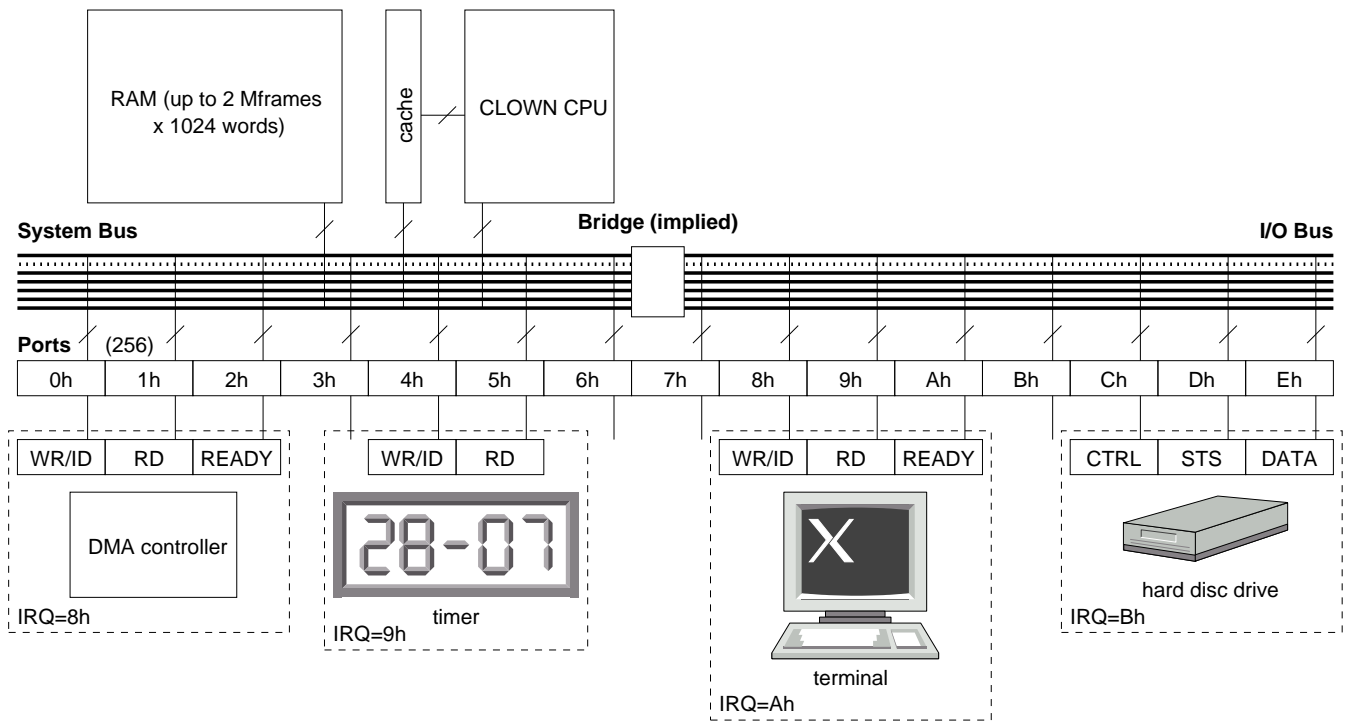, width=\textwidth}
\caption{\label{system}Clown system architecture}
\end{figure*}

A wish list for an OS-optimized simulator includes the following
requirements:

\begin{mylist}
\item Rich support for OS concepts.
\item Little or no support for application-specific features, such as
  string operations and floating-point unit (to reduce complexity and
  learning time).
\item Reasonably detailed simulation (to make sure that the simulator
  could also be used in a computer architecture course). 
\item A collection of basic I/O devices, with a mechanism for adding
  more devices, if needed.
\item Fast (preferably real-time) execution, ideally in real time
  emulation mode.
\item A simple interface.
\item A substantial set of development tools (such as assembler,
  linker, disk editor, debugger, C compiler).
\end{mylist}

To satisfy these requirements, I developed Clown --- a new simulator
of an Intel-style microprocessor and computer system specifically
tuned to the needs of the courses mentioned above.

\section{CLOWN OVERVIEW}

Clown simulator suite is partially based on the Simple Hard Disk
Emulator (SHaDE~\cite{shade02}) written at Suffolk University as a
simple vehicle for teaching the low-level organization of file
systems.

The system architecture of Clown is shown in Figure~\ref{system}.

The simulator consists of a Clown CPU with a single-level
direct-mapped write-back cache (included to simulate DMA transfers
accurately), one bank of 32-bit non-interleaved memory, 32-bit system
and I/O buses with an implied bridge (the bridge is not simulated, and
both buses are treated as one bus), 256 I/O ports (Intel has 65,536
ports), 16 interrupt channels, and one DMA channel (Intel typically has
7 DMA channels). Four basic I/O devices are included in the standard
configuration. 

\begin{figure}[tb!]\centering
\epsfig{file=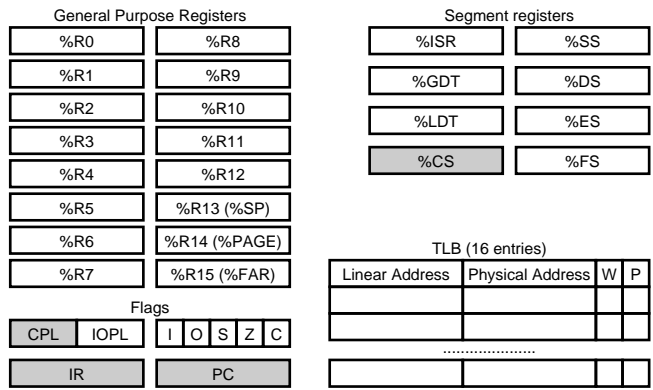, width=\columnwidth}
\caption{\label{cpu}Clown CPU architecture; shaded registers and flags
are not accessible from programs}
\end{figure}

The architecture of the Clown CPU is shown in Figure~\ref{cpu}.

The CPU has sixteen 32-bit general-purpose registers (Intel: 8 GPR and
two control registers, CR0 and CR3), eight 32-bit segment
registers\footnote{This number is redundant and can be reduced to
six.} (Intel: 6 segment and 4 memory management registers), one 16-bit
flag register, an instruction register and a program counter. A
16-entry direct-mapped Translation Look-aside Buffer contributes to
the accuracy of DMA transfers (Intel 80386 has a 32-entry 4-way
set-associative TLB). There is no dedicated stack pointer register,
page table base register, and page fault address register. Their
functions are assigned to general-purpose registers \%R13 throught
\%R15.

Clown supports only one data type: signed 32-bit word (for comparison,
Intel supports at least 12 data types~\cite{smith87}). This feature
drastically simplify system programming. On the other hand, it poses
interesting challenges to compiler developers (such as type
representations and conversions, and implementation of floating point
arithmetics).

The Clown CPU supports both paging and segmentation. Either memory
organization mechanism can be turned off (by disabling the page table
or by declaring all memory to be one big implicit segment).

In the Intel architecture, an interrupt vector (IV) is always treated
as an array of segment descriptors, each identifying an entry into an
interrupt service routine (ISR).  In pure paging mode, it would be
highly desirable to have no segments whatsoever, including the
ISRs. This is accomplished by forcing all ISRs to be 8-word aligned,
and treating the least significant bit of an IV entry as a mode
bit. When this bit is clear, the IV entry is treated as a segment
descriptor. If the bit is set, the entry is treated as the direct
address of the entry point followed by two protection bits. While this
approach does not seem elegant enough, it nevertheless allows the
development of segment-free operating systems.

\begin{table*}[tb!]\centering
\caption{\label{isa}Comparison of the Clown and i386 instruction sets}
\vskip0.5\baselineskip
\begin{tabular}{|l|c|c||l|c|c|}\hline
{\bf Group} & {\bf i386} & {\bf Clown} & {\bf Group} & {\bf i386} & {\bf Clown}\\\hline
Data movement & 8 & 13 & Arithmetic & 12 & 12\\
Shift / Rotate & 12 & 8 & Logical & 6 & 11\\
Bits and Bytes & 39 & 8 & Flag control &11&6\\
Processor control & 4 & 3&Flow control & 77 & 18\\
Memory protection & 1 & 5&I/O & 4 & 3\\
Other & 25 & 0&&&\\\hline
\multicolumn{4}{|l|}{\bf Total} & 200 & 87\\\hline
\end{tabular}
\end{table*}

Compared to i386, Clown has significantly fewer instructions, which
reduces the learning time (Table~\ref{isa}). Explicitly omitted are
data conversion instructions, decimal arithmetics, address
manipulation, string, and translation instructions, and high-level
language support instructions.

A Clown instructions consist of either one or two words. The second
word, if present (recognized by the MSB of the first word, or by the
``{\tt x}'' prefix in the mnemonics), is always the immediate operand.

The number of flags has also been minimized. There are only 7
externally visible flags: \underline{C}arry, \underline{Z}ero,
\underline{S}ign, \underline{O}verflow, \underline{I}nterrupts
(enabled), and two \underline{I}/\underline{O} \underline{P}rivilege
\underline{L}evel flags (compared to 13 flags in Intel 30386).

Clown runs a simple fetch-decode-execute loop. The execution of each
instruction takes exactly one Clown cycle. External interrupts are
reported and queued at the end of a cycle. Nested interrupts are
permitted, with high-precedence interrupts preempting low-precedence
interrupts. This simulation model may change in the future to better
reflect modern pipelined architectures and their impact on process
context switches.

\section{PERIPHERAL DEVICES}

Currently, Clown has four peripheral devices: interval timer,
terminal, hard disk controller, and direct memory access (DMA)
controller. Each device has a configurable I/O base and a configurable
IRQ channel. All devices can operate in both polling and interrupt
modes.

The interval timer works both in interval and single-shot modes. It
generates an interrupt upon expiration, and can be stopped at any
time. The following assembly code fragment programs the timer to
expire (once) in 1000 cycles:

\vskip0.5\baselineskip
\noindent%
{\tt\strut \#include "config.h"}\\
{\tt\strut\ \ \ \ \ \ \ ; reset timer}\\
{\tt\strut\ \ \ \ \ \ \ out 1, (IOBASE\_TIMER + 0)}\\
{\tt\strut\ \ \ \ \ \ \ ; set the counter and trigger timer}\\
{\tt\strut\ \ \ \ \ \ \ out 1000, (IOBASE\_TIMER + 0)}\\
{\tt\strut\ \ \ \ \ \ \ ; wait for an interrupt}\\
{\tt\strut\ \ \ \ \ \ \ hlt}
\vskip0.5\baselineskip

The terminal combines a keyboard and a sequentially accessible (not
memory-mapped) display. When in the interrupt mode, it generates
interrupts on keystrokes. The terminal does not echo characters
(echoing is left to the programmer).

The hard disk controller carefully simulates the mechanical behavior
of a relatively simple hard disk (including track-to-track and maximum
seek latency, and rotational latency). The dynamic parameters of the
disk are run-time configurable. Inter-sector gaps make it possible to
optimize file systems for high-speed streaming operations. When in the
interrupt mode, the controller generates interrupts on completion of
seek, read, and write operations. A one-block read-write buffer is
prehistorically tiny, but yet sufficient to study the foundations of
disk I/O subsystem. At most one I/O request can be pending at any
time, so no disk scheduling is provided.

\begin{figure}[b!]\centering
\epsfig{file=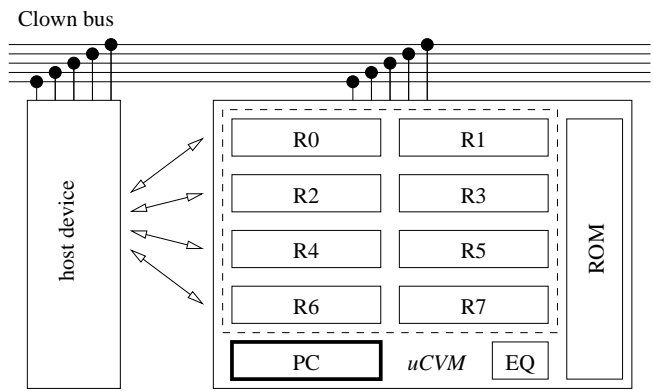, width=\columnwidth}
\caption{\label{ucvm}$\mu$CVM system architecture}
\end{figure}

The DMA controller is the most intelligent peripheral device. Clown
carefully simulates DMA transfers; data are transferred only when the
bus is not used by the main CPU. A transfer unit is fixed and equal to
one disk block (one virtual memory page). When in the interrupt mode,
the controller generates an interrupt on completion of the transfer
(which happens {\em after} the completion of the respective disk {\em
read} operation or {\em before} the completion of the respective disk
{\em write} operation).

The DMA controller works concurrently with the rest of the Clown
system. Its implementation as a part of the main fetch-decode-execute
loop would involve complex serialization and synchronization
issues. For instance, one Clown {\tt out} instruction triggers a
disk-to-memory transfer which takes a significant and uncertain number
of instructions to complete (due to seek and rotational
latencies). Calling a {\em read\_sector} function is not an option,
because it would hinder the main loop.

\begin{table*}[tb!]\centering
\caption{\label{ucvmisa}$\mu$CVM Instruction Set}
\vskip0.5\baselineskip
\begin{tabular}{|l|lll|l||l|lll|l|}\hline
Opcode&\multicolumn{3}{|c|}{Instruction}&Description&
Opcode&\multicolumn{3}{|c|}{Instruction}&Description\\\hline
\multicolumn{5}{|c||}{\bf Single-word instructions}&
\multicolumn{5}{|c|}{\bf Double-word instructions}\\\hline
\multicolumn{10}{|c|}{\bf Arithmetic and control instructions (AC)}\\\hline
0h & NOP & && Do nothing & 4h & xMOVI & reg & val & Store a constant\\
1h & JEQ & dest & & Conditional jump& 5h&xADDI & reg & val & Add a constant\\
2h & JMP & dest & & Unconditional jump& 6h&xCMPI & reg & val & Compare\\
3h & END &  & & Stop the VM&7h&&&&reserved\\\hline
\multicolumn{10}{|c|}{\bf I/O and memory instructions (IOM)}\\\hline
8h & OUT & port&reg&Output to the port&Ch&xOUTI&port&val&Output a constant\\
9h & IN & port&reg&Input from the port&Dh&&&&reserved\\
Ah & ST & [reg]&reg&Store indirectly&Eh&&&&reserved\\
Bh & LD & [reg]&reg&Load indirectly&Fh&&&&reserved\\\hline
\end{tabular}
\end{table*}

As a result, the controller is implemented using $\mu$CVM
(Microcontroller Virtual Machine) to enable true concurrent execution
of the main simulator and the simulator of the controller and also to
make the controller potentially reconfigurable. $\mu$CVM is a
``micro-Clown'': it has 8 general-purpose registers, a one-bit flag
register, and a program counter. The instruction set consists of 10
commands (see Table~\ref{ucvmisa}).

The main loop of the simulator first executes the next Clown
instruction. If it is not a memory reference or an I/O instruction,
then the next $\mu$CVM instruction is executed. Otherwise, the next
$\mu$CVM instruction is executed only if it is not a memory reference
or an I/O instruction. 

The main program of the $\mu$CVM that controls both disk-to-memory and
memory-to-disk transfers, fits in just 132 bytes of the controller
memory.

The code simulating the peripheral devices is organized as dynamically
loaded libraries (one device per library). This organization turned
out to be flawed: while it did not significantly contribute to the
reconfigurability of the simulator, it undermined its integrity. In
the future releases, all code pieces will be linked together.

\section{ASSEMBLY LANGUAGE}

Clown assembly ({\em cas}) language uses a mixture of ``Intel-style''
and ``MIPS-style'' syntax. It has 53 commands and 90
modifications. The language allows decimal, octal, and hex numbers (in
prefix and postfix notation), and ASCII characters and
strings. Because Clown does not have a byte data type, characters and
strings are translated into words and arrays of words, one character
per word. This cumbersome conversion leads to ``sparse'' strings and
poor memory utilization, but significantly reduces the number of
machine instructions.

Before assembling, the source code is run through a standard C
preprocessor ({\em cpp}).

The {\em cas} assembler supports multiple segments (if needed) and
global symbols. It can produce raw ``bin'' executable files, without
headers and symbol tables, and structured multi-segment ``exe'' files,
with symbol tables and provisions for further linking with other files
of the same kind, using the Clown linker ({\em clink}). ``exe'' files
are fully relocatable. The displacement of the entry point into a
``bin'' file can be specified at the assembly time. ``Bin'' files can
be used as directly loadable ROM/RAM images. Clown simulator can
simultaneously load several executable images (for instance, to
simulate several processes without writing an OS loader).

So far, Clown does not include a run-time loader. Loaders are heavily
OS-dependent, and should be written by OS developers.

\section{FEASIBILITY AND PERFORMANCE EVALUATION}
The Clown architecture is meant to be feasible in the sense that,
whether a need arises to implement it in either FPGA or directly in
hardware, it will not pose significant risks and challenges. This
estimation is based on the author's experience with the FLUX
superconductor microprocessor design~\cite{zinoviev00}.

In the experiments, the Clown system simulated 4 million instructions
per second (4~MIPS) on a 1.3~GHz Pentium host CPU (native performance
2600~MIPS). The following code was used for performance evaluation:

\vskip0.5\baselineskip
\noindent%
{\tt\strut\ \ \ \ \ \ \ mov \%r1, 10000000}\\
{\tt again:\ dec \%r1}\\
{\tt\strut\ \ \ \ \ \ \ jnz again}\\
{\tt\strut\ \ \ \ \ \ \ stop}
\vskip0.5\baselineskip

This performance is roughly equivalent to Intel 8086 (4.77~MHz), which
is reasonably good for real-time user interaction.

\section{CLOWN IN A CLASSROOM}

The Clown system was ``field tested'' in an undergraduate Operating
Systems course during the Spring 2004 semester. The following eight
assignments were given throughout the semester to the students who had
taken an introductory assembly language course based on Intel 8086
architecture. For each assignment, the typical length of the program
code in non-commented lines of code (NLOC) is given.

\begin{mylist}
\item {\tt kputs} --- display a character string, using the terminal;
60 NLOCs 
\item {\tt boot} --- load the first sector of the first track, using
  polling, and execute its contents; 25 NLOCs 
\item {\tt boot-dma} --- load the first sector of the first track,
  using DMA,  and execute its contents; 15 NLOCs 
\item {\tt int-timer} --- populate and test the interrupt vector (timer
  ISR); 60 NLOCs 
\item {\tt int-kbd} --- populate and test the interrupt vector
  (timer ISR and keyboard ISR, competing for a counter variable); 85 NLOCs 
\item {\tt page-table} --- populate and test the page table; 15 NLOCs 
\item {\tt page-fault} --- populate and test the interrupt vector
  (page fault handler); 25 NLOCs  
\item {\tt file} --- traverse a disk file organized as a linked list;
  30 NLOCs 
\end{mylist}

Out of 14 students taking the class, nine successfully completed all
assignments, three completed 7 assignments, and the remaining two
completed 5 assignments.

\section{CONCLUSION AND FUTURE WORK}

Clown system is a powerful, simple, fast, configurable, and extensible
microprocessor simulator which can be used in various college-level
courses, especially in those dealing with operating systems.

Future work includes developing a debugger, a collection of sample
run-time loaders for multi-segment executable files, a C compiler, and
a graphical user interface. The simulator may be further optimized for
speed. Pipelining support needs to be added to provide more realistic
simulation if the package is to be used in a computer architecture
class. Networking, mouse, and graphics mode support would enable many
other uses of the simulator (such as a vehicle in a PDA graphics
study).

\section{ACKNOWLEDGMENTS}

I would like to thank Professors D.~Stefan\u{a}scu, D.~Cohn, and
A.~Thomo of Suffolk University for their encouraging support for the
project, Mrs.~P.~Salla for the partial implementation of the first
SHaDE emulator, and all students of my Spring 2004 Operating Systems
class for taking on the burden and pain of testing the package.

\section{BIOGRAPHY}

Dr. D.~Zinoviev received his Ph.D. in Computer Science from SUNY at
Stony Brook in 1997. He was working as a post-doc on the
DARPA/NASA/NSA-sponsored Petaflops project of a hybrid technology,
multi-threaded hypercomputer. In 2000, he joined the Computer Science
Department of Suffolk University in the rank of Assistant
Professor. His current research interests include simulation and
modeling (network simulation, architectural simulation), operating
systems, and software engineering (complexity metrics).

\bibliographystyle{plain}
\bibliography{biblio}

\end{document}